\documentclass[aps,superscriptaddress,showpacs,showkeys]{revtex4}

\usepackage{graphicx}
\usepackage[usenames]{color}
\usepackage{lscape,fancybox,pifont}
\usepackage{wasysym}

\begin{document}

\title{Collective dynamics of social annotation}

\author{Ciro Cattuto}
\affiliation{Complex Systems and Networks
 Lagrange Laboratory (CNLL), ISI Foundation, Turin, Italy}
\author{Alain Barrat}
\affiliation{Centre de Physique Th\'eorique (CNRS UMR 6207), 
Campus de Luminy, Case 907, 13288 Marseille cedex 9, France}
\affiliation{Complex Systems and Networks
 Lagrange Laboratory (CNLL), ISI Foundation, Turin, Italy}
\author{Andrea Baldassarri}
\affiliation{Dipartimento di Fisica, ``Sapienza'' Universit\`a di
Roma, Piazzale Aldo Moro 5, 00185 Roma, Italy}
\author{Gregory Schehr}
\affiliation{Laboratoire de Physique Th\'eorique (CNRS UMR8627),
Universit\'e Paris-Sud, 91405 Orsay cedex, France}
\author{Vittorio Loreto}
\affiliation{Dipartimento di Fisica, ``Sapienza'' Universit\`a di
Roma, Piazzale Aldo Moro 5, 00185 Roma, Italy}
\affiliation{Complex Systems and Networks
 Lagrange Laboratory (CNLL), ISI Foundation, Turin, Italy}

\begin{abstract}
The enormous increase of popularity and use of the WWW has led in the
recent years to important changes in the ways people communicate. An
interesting example of this fact is provided by the now very popular
social annotation systems, through which users annotate resources
(such as web pages or digital photographs) with text keywords dubbed
tags. Understanding the rich emerging structures resulting from the
uncoordinated actions of users calls for an interdisciplinary
effort. In particular concepts borrowed from statistical physics, such
as random walks, and the complex networks framework, can effectively
contribute to the mathematical modeling of social annotation
systems. Here we show that the process of social annotation can be
seen as a collective but uncoordinated exploration of an underlying
semantic space, pictured as a graph, through a series of random
walks. This modeling framework reproduces several aspects, so far
unexplained, of social annotation, among which the peculiar growth of
the size of the vocabulary used by the community and its complex
network structure that represents an externalization of semantic
structures grounded in cognition and typically hard to access.
\end{abstract}

\maketitle


\section{Introduction}

The rise of Web 2.0 has dramatically changed the way in which
information is stored and accessed, and the relationship between
information and on-line users. This is prompting the need for a new
research agenda about ``Web Science'', as put forward
in~\cite{webscience}. A central role is played by user-driven
information networks, i.e., networks of on-line resources built in a
bottom-up fashion by Web users. These networks entangle cognitive,
behavioral and social aspects of human agents with the structure of
the underlying technological system, effectively creating
techno-social systems that display rich emergent features and
\textit{emergent semantics}~\cite{staab2002_emergent,mika2007}.
Understanding their structure and evolution brings forth new
challenges.

Many popular Web applications are now exploiting user-driven
information networks built by means of \textit{social
  annotations}~\cite{wu2006_exploring,hhls05social}.  Social
annotations are freely established associations between Web resources
and metadata (keywords, categories, ratings) performed by a community
of Web users with little or no central coordination.  A mechanism of
this kind which has swiftly become well-established is that of
\text{collaborative
  tagging}~\cite{mathes04folksonomies,golder2006structure}, whereby
Web users associate free-form keywords -- called ``tags'' -- with
on-line content such as Web pages, digital photographs, bibliographic
references and other media.
The product of the users' tagging activity
is an open-ended information network -- commonly referred to
as ``folksonomy'' -- which can be used for navigation and recommendation
of content, and has been the object of many recent investigations
across different disciplines~\cite{marlow2006_ht06,cattuto2007pnas}.
Here we show how simple concepts borrowed from statistical physics and
the study of complex networks can provide a modeling framework for the
dynamics of collaborative tagging and the structure of the 
ensuing folksonomy.

Two main aspects of the social annotation process, so far unexplained,
deserve a special attention. One striking feature is the so-called
Heaps' law~\cite{heaps78} (also known as Herdan's law in linguistics),
originally studied in Information Retrieval for its relevance for
indexing schemes~\cite{baezayates00block}. Heaps' law is an empirical
law which describes the growth in a text of the number of
distinct words as a function of the number of total words
scanned. It describes thus the rate of innovation
in a stream of words, where innovation means the adoption for
the first time in the text of a given word. This law, also
experimentally observed in streams of tags, consists of a
power-law with a sub-linear
behavior~\cite{cattuto2007pnas,cattuto2007vocabulary}. In this case
the rate of innovation is the rate of introduction of new
tags, and a sub-linear behavior 
corresponds to a rate of adoption of new words or tags decreasing with
the total number of words (or tags) scanned. Most existing
studies about Heaps' law, either in Information Retrieval or in
linguistics, explained it as a consequence of the so-called Zipf's
law~\cite{zipf49,baezayates00block}. It would instead be
highly desirable to have an explanation for it relying only on very
basic assumptions on the mechanisms behind social annotation.

Another important way to analyze the emerging data structures is given
by the framework of complex
networks~\cite{dorogovtsev03,Pastor:2004,Barrat:2008}. These
structures are indeed user-driven information networks
\cite{cattuto2007aicomm}, i.e., networks linking (for instance)
on-line resources, tags and users, built in a bottom-up fashion
through the uncoordinated activity of thousands to millions of Web
users.  We shall focus in particular on the particular structure of
the so-called co-occurrence network. The co-occurrence network is a
weighted network where nodes are tags and two tags are linked if they
were used together by at least one user, the weight being larger when
this simultaneous use is shared by many users. Correlations between
tag occurrences are (at least partially) an externalization of the
relations between the corresponding
meanings~\cite{sowa84,sole08languagenetworks} and have been used to
infer formal representations of knowledge from social
annotations~\cite{heymann06collaborative}. Notice that co-occurrence
of two tags is not a priori equivalent to a semantic link between the
meanings/concepts associated with those tags, and that understanding
what co-occurrence precisely means, in terms of semantic relations of
the co-occurring tags, is an open question that is investigated in
more applied contexts \cite{cattuto2008iswc,cattuto08-semantic}.

On these aspects of social annotation systems,
a certain number of stylized facts, about e.g. tag
frequencies~\cite{golder2006structure,cattuto2007pnas} or the growth
of the tag vocabulary~\cite{cattuto2007vocabulary}, have been reported
but no modeling framework exists which can naturally account for them
while reproducing the co-occurrence network structure. Here we ask
whether one is able to explain the structure of such a network in
terms of some suitable generative model and how the structure of the
experimentally observed co-occurrence network is related to the
underlying hypotheses of the modeling scheme. We show in particular
that the idea of social exploration of a semantic space has more than
a metaphorical value, and actually allows us to reproduce
simultaneously a set of independent correlations and fine observables
of tag co-occurrence networks as well as robust stylized facts of
collaborative tagging systems.

\section{User-driven information networks}

We investigate user-driven information networks using data from two
social bookmarking systems:
del.icio.us\footnote{{http://del.icio.us}} and
BibSonomy\footnote{{http://www.bibsonomy.org}}. Del.icio.us is a
very popular system for bookmarking web pages and pioneered the
mechanisms of collaborative tagging.  It hosts a large body of social
annotations that have been used for several scientific
investigations. BibSonomy is a smaller system for bookmarking
bibliographic references and web pages~\cite{hjss06bibsonomy}.  Both
del.icio.us and BibSonomy are \emph{broad}
folksonomies~\cite{vanderwal}, in which users provide metadata about
pre-existing resources and multiple annotations are possible for the
same resource, making the ensuing tagging patterns truly ``social''
and allowing their statistical characterization. 

A single user annotation, also known as a \emph{post}, is a triple of
the form $(u, r, T)$, where $u$ is a user identificator, $r$ is the
unique identificator of a resource (a URL pointing to a web page, for
the systems under study), and $T = \{t_1, t_2, \dots \}$ is a set of
tags represented as text strings. We define the tag co-occurrence
network based on post co-occurrence. That is, given a set of posts, we
create an undirected and weighted network where nodes are tags and two
tags $t_1$ and $t_2$ are connected by an edge if and only if there
exists one post in which they were used in conjunction. The weight
$w_{t_1 t_2}$ of an edge between tags $t_1$ and $t_2$ can be naturally
defined as the number of distinct posts where $t_1$ and $t_2$
co-occur. This construction reflects the existence of semantic
correlations between tags, and translates the fact that these
correlations are stronger between tags co-occurring more frequently.
We emphasize once again that the co-occurrence network is an
externalization of hidden semantic links, and therefore distinct from
underlying semantic lexicons or networks.

\subsection{Data from del.icio.us}
The del.icio.us dataset we used consists of approximately $5 \cdot
10^6$ posts, comprising about $650\,000$ users, $1.9 \cdot 10^6$
resources (bookmarks) and $2.5 \cdot 10^6$ distinct tags. It covers
almost $3$ years of user activity, from early 2004 up to November
2006.  Overall, $667\,128$ user pages of the del.icio.us community
were crawled, for a total of $18\,782\,132$ resources, $2\,454\,546$
distinct tags, and $140\,333\,714$ tag assignments (triples).

The data were subsequently post-processed for the present study.  We
discarded all posts containing no tags (about $7$\% of the total).  As
del.icio.us is case-preserving but not case sensitive, we ignored
capitalization in tag comparison, and counted all different
capitalizations of a given tag as instances of the same lower-case
tag.  The timestamp of each post was used to establish post ordering
and determine the temporal evolution of the system.  Posts with
invalid timestamps, i.e.\ times set in the future or before
del.icio.us started operating, were discarded as well (less than
$0.5$\% of the total).

Except for the normalization of character case, no lexical
normalization was applied to tags during post-processing. The notion
of identity of tags is identified with the notion of identity of their
string representation.

\subsection{Data from BibSonomy}
BibSonomy~\cite{hjss06bibsonomy} is a smaller system than del.icio.us,
but it was designed keeping data sharing in mind. Because of this,
there is no need to crawl BibSonomy by downloading HTML pages and
parsing them.  Direct access to post data in structured form is
available by using the BibSonomy API
({http://www.bibsonomy.org/help/doc/api.html}).  Moreover, the
BibSonomy team periodically releases snapshot datasets of the full
system and makes them available to the research community.  For the
present work we used the dataset released on January 2008
({https://www.kde.cs.uni-kassel.de/bibsonomy/dumps/2007-12-31.tgz}).

BibSonomy allows two different types of resources: bookmarks (i.e.,
URLs of web pages, similar to del.icio.us) and BibTeX entries. To make
contact with the analysis done for del.icio.us, we restricted the
dataset to the posts involving bookmark resources only.  The resulting
dataset we used comprises $1\,400$ users, $127\,115$ resources,
$37\,966$ distinct tags, and $503\,928$ tag assignments (triples).
The data from BibSonomy was post-processed in the same way as the data
from del.icio.us.

While the BibSonomy dataset is much smaller than the del.icio.us
dataset, it is a precious one: direct access to BibSonomy's database
guarantees that the BibSonomy dataset is free from biases due to the
data collection procedure. This is important because it allows us to
show that the investigated features of the data are robust across
different systems, and not only established in a case where biases due
to data collection could be possible.

\subsection{Data analysis}

\begin{figure}[htb]
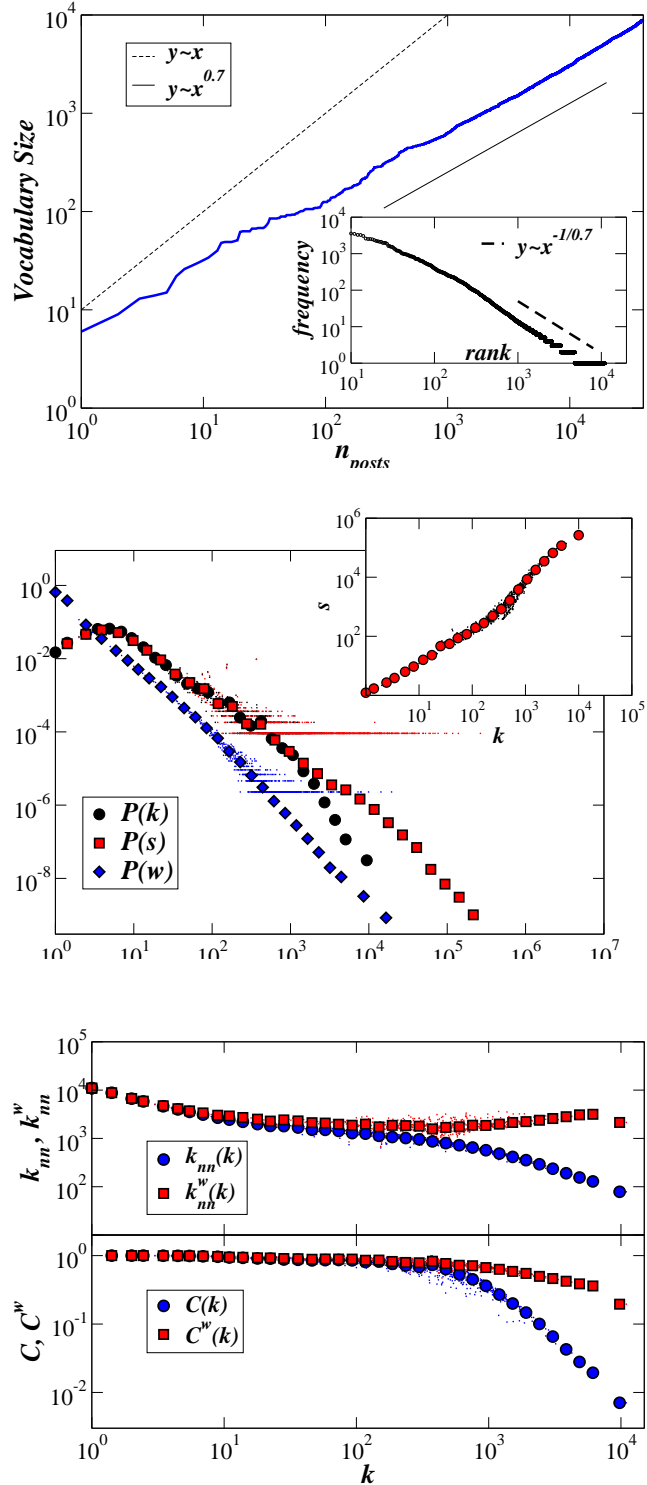

\begin{center}
\vspace{-0.5cm}
\includegraphics[width=0.47\textwidth]{fig_1a}
\vskip0.4cm
\includegraphics[width=0.47\textwidth]{fig_1b}
\vskip0.4cm
\includegraphics[width=0.47\textwidth]{fig_1c}
\caption{ Data corresponding to the posts containing the
  tag "Folksonomy" in del.icio.us.  Top: Heaps' law: growth of the
  vocabulary size associated with the tag $t^*=$"Folksonomy", measured
  as the number of distinct tags co-occurring with $t^*$, as a
  function of the number $n_{posts}$ of posts containing $t^*$.  The
  dotted line corresponds to a linear growth law while the continuous
  line is a power-law growth with exponent $0.7$.  Inset: Frequency-rank
  plot of the tags. The dashed line corresponds to a power-law $-1.42
  \simeq -1./0.7$.  Middle and Bottom: Main properties of the co-occurrence
  network of the tags co-occurring with the tag "Folksonomy" in
  del.icio.us, built as described in the main text. Middle figure: Broad
  distributions of degrees $k$, strengths $s$ and weights $w$ are
  observed. The inset shows the average strength of nodes of degree
  $k$, with a superlinear growth at large $k$. Bottom figure: Weighted
  ($k_{nn}^w$) and unweighted ($k_{nn}$) average degree of nearest
  neighbors (top), and weighted ($C^w$) and unweighted ($C$) average
  clustering coefficients of nodes of degree $k$.  $k_{nn}$ displays a
  disassortative trend, and a strong clustering is observed.  At small
  $k$, the weights are close to $1$ ($s(k)\sim k$, see inset of middle figure,
  and $k_{nn}^w\sim k_{nn}$, $C^w\sim C$. At large $k$ instead,
  $k_{nn}^w > k_{nn}$ and $C^w > C$, showing that large weights are
  preferentially connecting nodes with large degree: large degree
  nodes are joined by links of large weight, i.e. they co-occur
  frequently together. In (B) and (C) both raw and logarithmically
  binned data are shown.}
\label{fig1}
\end{center}
\end{figure}

The study of the global properties of the tagging system, and in
particular of the global co-occurrence network, is of interest but
mixes potentially many different phenomena. We therefore consider a
narrower semantic context, defined as the set of posts containing one
given tag.  We define the vocabulary associated with a given tag $t^*$
as the set of all tags occurring in a post together with $t^*$, and
the time is counted as the number of posts in which $t^*$ has
appeared. The size of the vocabulary follows a sub-linear power-law
growth (Fig.~\ref{fig1}), similar to the Heaps' law~\cite{heaps78}
observed for the vocabulary associated with a given resource, and for
the global vocabulary \cite{cattuto2007vocabulary}. 
Figure~\ref{fig1} also displays the main properties of the
co-occurrence network, as measured by the quantities customarily used
to characterize statistically complex networks and to validate
models~\cite{Pastor:2004,Barrat:2008}. These quantities can be
separated in two groups.  On the one hand, they include the
distributions of single node or single link quantities, whose
investigations allow to distinguish between homogeneous and
heterogeneous systems. Figure~\ref{fig1} shows that the co-occurrence
networks display broad distributions of node degrees $k_t$ (number of
neighbors of node $t$), node strengths $s_t$ (sum of the weights of
the links connected to $t$, $s_t= \sum_{t'} w_{t t'}$), and link
weights. The average strength $s(k)$ of vertices with degree $k$,
$s(k)=\frac{1}{N_k}\sum_{t/k_t=k} s_t$, where $N_k$ is the number of
nodes of degree $k$, also shows that correlations between topological
information and weights are present. On the other hand, these
distributions by themselves are not sufficient to fully characterize a
network and higher order correlations have to be investigated. In
particular, the average nearest neighbors degree of a vertex $t$,
$k_{nn,t}=\frac{1}{k_t}\sum_{t' \in {\cal V}(t)} k_{t'}$, where ${\cal
  V}(t)$ is the set of $t$'s neighbors, gives information on
correlations between the degrees of neighboring nodes.
Moreover, the clustering coefficient $c_t= e_t/(k_t(k_t-1)/2)$ of a
node $t$ measures local cohesiveness through the ratio between the
number $e_t$ of links between the $k_t$ neighbors of $t$ and the
maximum number of such links \cite{Watts:1998}. The functions
$k_{nn}(k) = \frac{1}{N_k} \sum_{t/k_t=k} k_{nn,t}$ and
$C(k)=\frac{1}{N_k}\sum_{t/k_t=k} c_t$ are convenient summaries of
these quantities, that can also be generalized to include weights (see
SI for the definitions of $k^w_{nn}(k)$ and $C^w(k)$).
Figure~\ref{fig1} shows that broad distributions and non-trivial
correlations are observed. All the measured features are robust
across tags within one tagging system, and also across the tagging
systems we investigated.

\section{Modeling social annotation}
The observed features are emergent characteristics of the
uncoordinated action of a user community, which call for a
rationalization and for a modeling framework. We now present a simple
mechanism able to reproduce the complex evolution and structure of the
empirical data.

\begin{figure}[htb]
\includegraphics[width=0.59\textwidth]{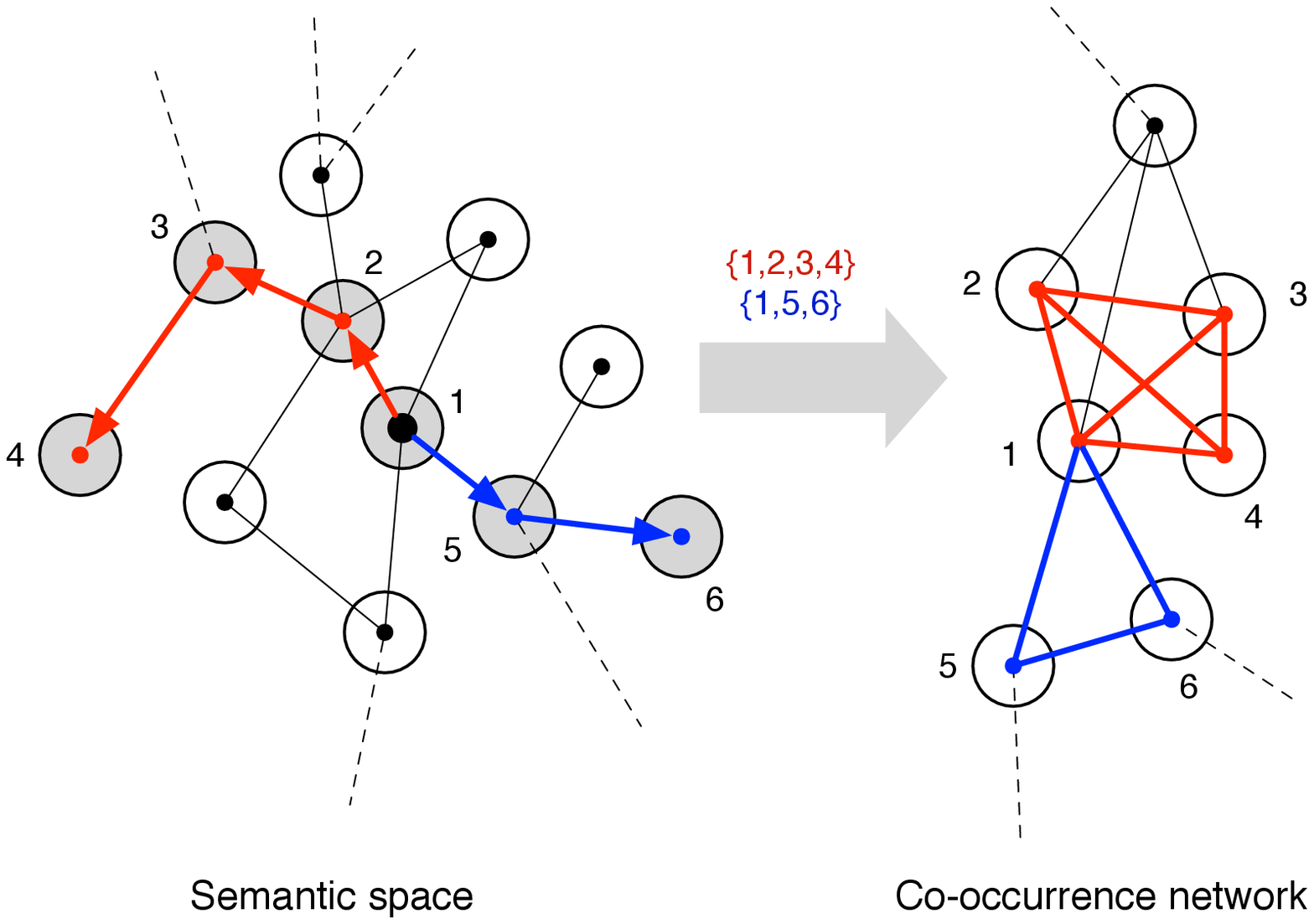}
\quad
\includegraphics[width=0.35\textwidth]{fig_2b}\\
\caption{ Left: Illustration of the proposed mechanism of social
  annotation. The semantic space is pictured as a network in which
  nodes represent tags and a link corresponds to the possibility of a
  semantic association between tags.  A post is then represented as a
  random walk on the network. Successive random walks starting from
  the same node allow the exploration of the network associated with a
  tag (here pictured as node $1$). The artificial co-occurrence
  network is built by creating a clique between all nodes visited by a
  random walk.  Right: empirical distribution of posts' lengths
  $P(l)$. A power-law decay $\sim l^{-3}$ (dashed line) is observed.}
\label{fig2}
\vskip0.4cm
\end{figure}

The fundamental idea underlying our approach, illustrated in
Fig.~\ref{fig2}, is that a post corresponds to a random walk (RW) of
the user in a ``semantic space'' modeled as a graph. Starting from a
given tag, the user adds other tags, going from one tag to another by
semantic association. It is then natural to picture the semantic space
as network-like, with nodes representing tags and links representing
the possibility of a semantic link \cite{Steyvers:2005}. A precise and
complete description of such a semantic network being out of reach, we
make very general hypothesis about its structure and we have checked
the robustness of our results with respect to different plausible
choices of the graph structure \cite{Steyvers:2005}.
Nevertheless, as we shall see later on, our results help fixing some
constraints on the structural properties of such a semantic space:
it should have a finite
average degree together with a small graph diameter, which ensures
that RWs starting from a fixed node and of limited length can 
potentially reach all nodes of the graph.
In this framework, the vocabulary co-occurring with a tag is associated
with the ensemble of nodes reached by successive random walks starting
from a given node, and its size with the number of {\em distinct}
visited nodes, $N_{distinct}$, which grows as a function of the number
of performed random walks $n_{RW}$. 

\subsection{Fixed length random walks}
Let us first consider random walks of fixed length $l$ starting from a 
given node $i_0$. We denote by $p_i$ the probability for each of these
random walks to visit node $i$. The probability that $i$ has \emph{not}
been visited after $n_{RW}$ random walks is then simply
\begin{equation}
\mbox{Proba(i not visited)}=(1-p_i)^{n_{RW}} \ ,
\end{equation}
since the random walks are independent stochastic processes,
and the probability that $i$ has been visited at least once reads
\begin{equation}
\mbox{Proba(i visited)=1 - Proba(i not visited)}=1 - (1-p_i)^{n_{RW}} \ .
\end{equation}
The average number of distinct nodes visited after $n_{RW}$ random walks
is then given, without any assumption on the network's structure, by
\begin{equation}
N_{distinct} = \sum_i \left( 1 - (1-p_i)^{n_{RW}} \right) \ ,
\end{equation}
where the sum runs over all nodes of the network.

While this exact expression is not yet really informative, it is
possible to go further under some simple assumptions (we also note
that analytical results are available in the case of random walks
performed either on lattices or on fractal substrates
\cite{Acedo:2005}). Since all the random walks start from the same
origin $i_0$, it is useful to divide the network into successive
``rings'' \cite{Baronchelli:2006}, each ring of label $l$ being formed
by the nodes at distance $l$ from $i_0$. The ring $l=1$ is formed by
the neighbours of $i_0$, the ring $l=2$ by the neighbours' neighbours
which are not part of ring $1$, and so forth. We denote by $N_l$ the
number of nodes in ring $l$.  We now make the assumption that all
$N_l$ nodes at distance $l$ have the same probability to be reached by
a random walk starting from $i_0$ (which is the sole element of ring
$0$). This is rigorously true for example for a tree with constant
coordination number, and more generally will hold approximately in
homogeneous networks, while stronger deviations are expected in
heterogeneous networks. Let us assume moreover that the random walk of
length $l_{max}$ consists, at each step, of moving from one ring $l$
to the next ring $l+1$. This is once again rigorously true for a
self-avoiding random-walk on a tree, and can be expected to hold
approximately if $N_l$ grows fast enough with $l$: the probability to
go from ring $l$ to ring $l+1$ is then larger than to go back to ring
$l-1$ or to stay within ring $l$. For each random walk of length
$l_{max}$, we then have $p_i=1/N_l$ for each node $i$ in ring $l \le
l_{max}$, and after $n_{RW}$ walks, the average number of distinct
visited nodes reads
\begin{equation}
N_{distinct} = \sum_{l=0}^{l_{max}} N_l ( 1 - (1-1/N_l)^{n_{RW}}) \ .
\label{eqNl}
\end{equation}

The expression (\ref{eqNl}) lends itself to numerical investigation
using various forms for the growth of $N_l$ as a function of $l$.  We
obtain (not shown) that, as $n_{RW}$ increases, $N_{distinct}$ increases, with an
approximate power-law form, and saturates as $n_{RW} \to \infty$ at
the total number of reachable nodes $\sum_{l=0}^{l_{max}}
N_l$. Moreover, the increase at low $n_{RW}$ is sub-linear if $N_l$
grows fast enough with $l$ (at least $\sim l^2$), and is closer to
linear if $l_{max}$ increases.

\subsection{Random walks of randomlengths}
Empirical evidence on the distribution of post
lengths (Fig.~\ref{fig2}) suggests to consider random walks of random
lengths, distributed according to a broad law. 
Let us therefore now consider, under the same assumptions, that the successive
random walks have randomly distributed lengths according to a certain
$P(l)$. Each ring $l$, on average, is then reached by a random walk
$n_{RW} \times \sum_{l'\ge l} P(l') \equiv n_{RW} P_>(l)$ times, so
that we have approximately
\begin{equation}
N_{distinct} = \sum_{l=0}^{\infty} 
N_l ( 1 - (1-1/N_l)^{n_{RW}P_>(l)}) \ ,
\label{eqNlPl}
\end{equation}
where the sum (provided it converges) now runs over all possible lengths.

If $P(l)$ is narrowly distributed around an average value,
the form (\ref{eqNlPl}) will not differ very much from the case
of fixed length given by Eq. (\ref{eqNl}).
Conversely, for a broad $P(l)$, longer random walks will occur as $n_{RW}$
increases and the tail of $P(l)$ is sampled, allowing visits to nodes situated
further from $i_0$ and avoiding the
saturation effect observed for random walks of fixed length.

In some particular cases, a further analytical insight into the form
of $N_{distinct}(n_{RW})$ can be obtained:
\begin{itemize}
\item assume that $N_l \sim l^a$, and that $P(l)$ is power-law
  distributed ($P(l)\sim 1/l^b$). Then
  $N_{distinct}(n_{RW}) \sim \sum_{l=0}^\infty l^a
  \left(1- \exp(-n_{RW}/(c l^{a+b-1})) \right)$, where $c$ is a
  constant.  The terms in the sum become negligible for $l$ larger
  than $n_{RW}^{1/(a+b-1)}$, while they are close to $l^a$ for smaller
  values of $l$. The sum therefore behaves as
\begin{equation}
N_{distinct} \sim n_{RW}^{(a+1)/(a+b-1)} ,
\end{equation} 
i.e. a power-law.  For instance, for $b=3$
  we obtain a sub-linear power-law growth with exponent $(a+1)/(a+2)$,
  i.e. $2/3$ for $a=1$, or $3/4$ for $a=2$.

\item assume that $N_l \sim z^l$, which corresponds to a tree in which each
  node has $z+1$ neighbours, and $P(l) \sim 1/l^b$. Then 
$N_{distinct}(n_{RW}) \sim \sum_{l=0}^\infty z^l (1- \exp(-n_{RW}/(c z^l
  l^{b- 1}))$. As in the previous case, the terms in the sum become
  negligible for $l$ larger than $(log(n_{RW}) - (b-1)
  \log(\log(n_{RW}/\log(z))))/\log(z)$, 
while they are close to $z^l$ for smaller
  $l$. Thus the sum behaves as 
\begin{equation}
N_{distinct} \sim n_{RW}/(\log(n_{RW}))^{b-1}\ ,
\end{equation}  
i.e. we
  obtain a linear behaviour with logarithmic corrections, which is
  known to be very similar to sub-linear power-law behaviours.
\end{itemize}

We have thus shown analytically, under reasonable assumptions,
that performing fixed length random walks starting from
the same node yields a growth of the number
of distinct visited sites (representing the vocabulary size)
as a function of the number of random walks (representing posts)
which is sub-linear with a saturation effect,
and that broad distributions of the walks lengths lead to
sub-linear growths of the vocabulary, and avoid the saturation effect.

Figure \ref{fig3}(top) shows a confirmation of the appearance of a
sub-linear power-law-like growth of $N_{distinct}$, mimicking the
Heaps' law observed in tagging systems, for random walks performed
on a Watts-Strogatz network.

\section{Synthetic co-occurrence networks}

\begin{figure}[htb]
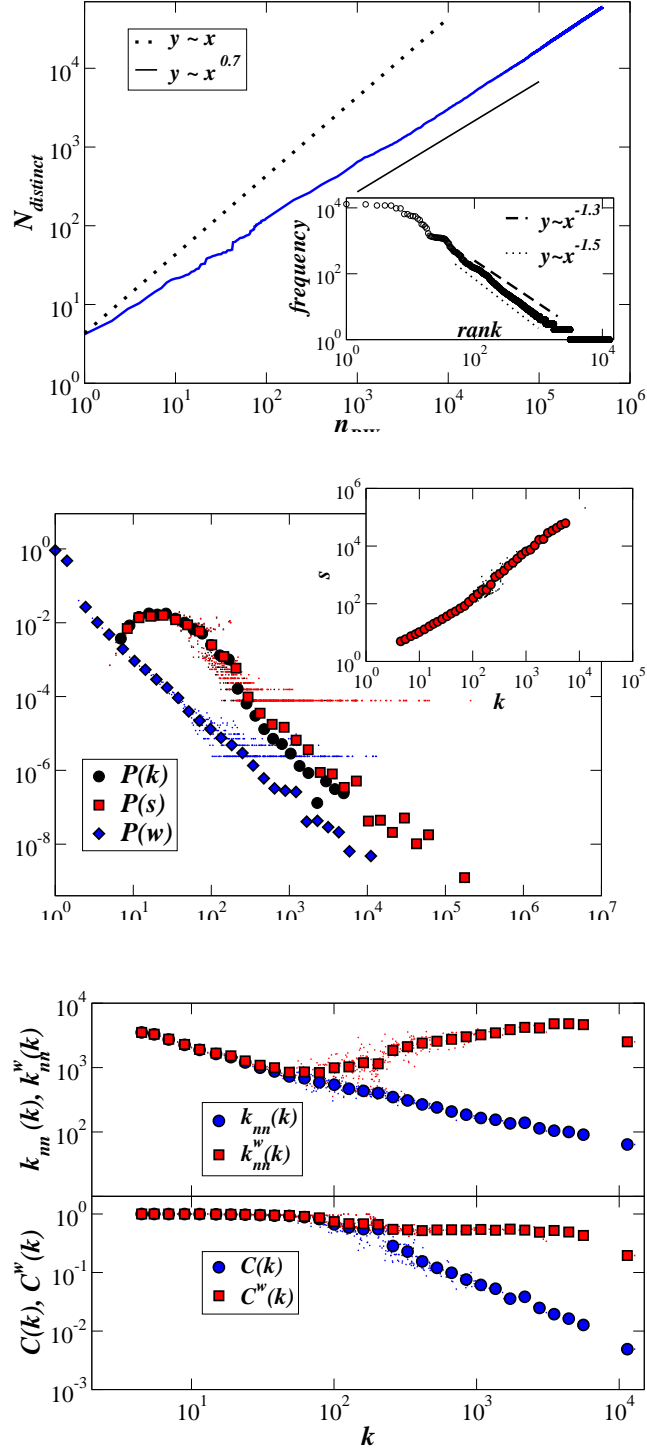

\includegraphics[width=0.47\textwidth]{fig_3a}
\vskip0.4cm
\includegraphics[width=0.47\textwidth]{fig_3b}\\
\vskip0.4cm
\includegraphics[width=0.47\textwidth]{fig_3c}
\caption{ Synthetic data produced through the proposed mechanism.
  Top: Growth of the number of distinct visited sites as a function of
  the number of random walks performed on a Watts-Strogatz network
  of size $5\cdot 10^4$ nodes and
  average degree $8$, rewiring probability $p=0.1$. Each random walk
  has a random length $l$ taken from a distribution $P(l) \sim
  l^{-3}$. The dotted line corresponds to a linear growth law while
  the continuous line is a power-law growth with exponent $0.7$. Inset:
  Frequency-rank plot. The continuous and dashed line have slope
  $-1.3$ and $-1.5$, respectively.  Middle and bottom: Properties of the
  synthetic co-occurrence network obtained for $n_{RW}=5\cdot 10^4$,
  to be compared with the empirical data of Fig.~\ref{fig1}.}
\label{fig3}
\end{figure}

\begin{figure}[htb]
\includegraphics[width=0.47\textwidth]{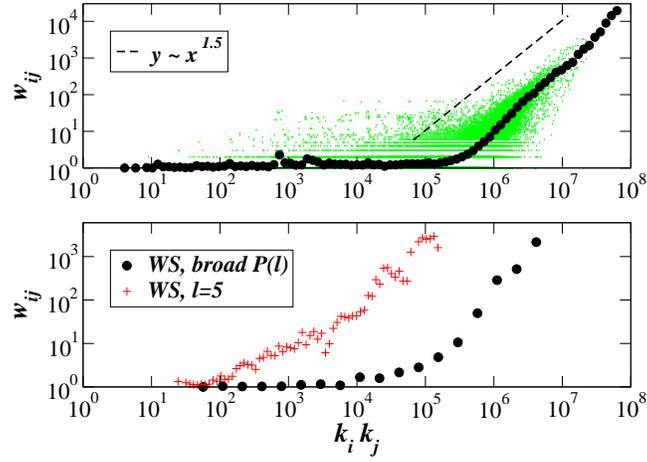}
\quad
\caption{Correlations between the weights of the links in the
  co-occurrence networks and the degrees of the links' endpoints, as
  measured by plotting the weight $w_{ij}$ of a link $i,j$ versus the
  product of the degrees $k_i k_j$.  Top: co-occurrence network of
  the tag "Folksonomy" of del.icio.us; each green dot corresponds to a
  link; the black circles represent the average over all links $i,j$
  with given product $k_i k_j$.  Bottom: synthetic co-occurrence
  networks obtained from $n_{RW}=5\cdot 10^4$ random walks performed
  on a Watts-Strogatz network of $10^5$ nodes. The black circles
  correspond to random walks of random lengths distributed according
  to $P(l)\sim l^{-3}$, and the red crosses to fixed length random
  walks ($l=5$).}
\label{fig4}
\vskip0.4cm
\end{figure}

Vocabulary growth is only one aspect of the dynamics of tagging
systems. Networks of co-occurrence carry much more detailed signatures
that present very specific features (Fig.~\ref{fig1}).  Interestingly,
our approach allows to construct {\em synthetic} co-occurrence
networks: we associate to each random walk a clique formed by the
nodes visited (see Fig.~\ref{fig2}), and consider the union of the
$n_{RW}$ such cliques. Moreover, each link $i,j$ built in this way
receives a weight equal to the number of times nodes $i$ and $j$
appear together in a random walk. This construction mimics precisely
the obtention of the empirical co-occurrence network, and also
reflects the idea that tags that are far apart in the underlying
semantic network are visited together less often than tags which are
semantically closer. Figures~\ref{fig3} and~\ref{fig4} show how the
  synthetic networks reproduce {\em all} statistical characteristics
  of the empirical data (Fig.~\ref{fig1}), both topological and
  weighted, including highly non-trivial correlations between topology
  and weights. Figure~\ref{fig4} in particular explores how the
  weight $w_{ij}$ of a link is correlated with its extremities'
  degrees $k_i$ and $k_j$. The peculiar shape of the curve can be
  understood within our framework. First, the broad distribution in
  $l$ is responsible for the plateau $\sim 1$ at small values of $k_i
  k_j$, since it corresponds to long RWs that occur rarely and visit
  nodes that will be typically reached a very small number of times
  (hence small weights). Moreover, $w_{ij} \sim (k_i k_j)^a$ at large
  weights. Denoting by $f_i$ the number of times node $i$ is visited,
  $w_{ij}\sim f_i f_j$ in a mean-field approximation that neglects
  correlations. On the other hand, $k_i$ is by definition the number
  of distinct nodes visited together with node $i$. Restricting the
  random walks to the only processes that visit $i$, it is reasonable
  to assume that such sampling preserves Heaps' law, so that $k_i
  \propto f_i^\alpha$, where $\alpha$ is the growth exponent for the
  global process. This leads to $w_{ij} \sim (k_i k_j)^a$ with
  $a=1/\alpha$. Since $\alpha \simeq .7-.8$, we obtain $a$ close to
  $1.3-1.5$, consistently with the numerics.

\begin{figure}[htb]
\includegraphics[width=0.47\textwidth]{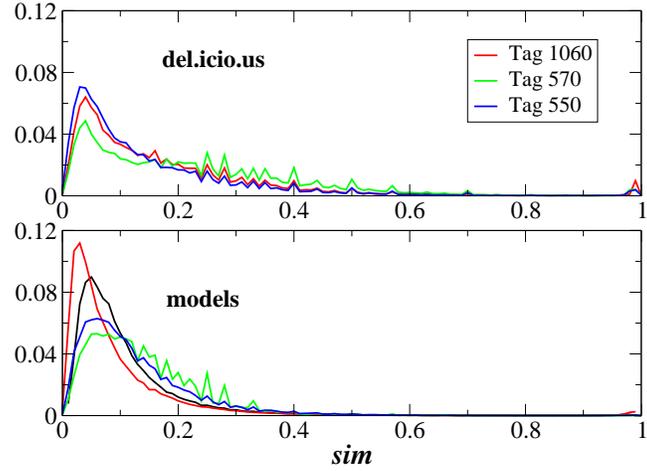}
\quad
\caption{Distributions of cosine similarities for real (top) and
  synthetic (bottom) co-occurrence networks. For del.icio.us, the tag
  number represents its popularity rank in the database. For
  the synthetic co-occurrence networks, the different curves correspond
  to different underlying networks on which the random walks are performed.}
\label{fig5}
\vskip0.4cm
\end{figure}

Strikingly, the synthetic co-occurrence networks reproduce
other, more subtle
observables, such as the distribution of cosine similarities between nodes.
In a weighted network, the similarity of two nodes $i_1$ and $i_2$ can
be defined as
\begin{equation}
\mbox{sim}(i_1,i_2) \equiv \sum_j
\frac{w_{i_1 j} w_{i_2 j}}
{\sqrt{\sum_\ell w_{i_1 \ell}^2 \sum_\ell w_{i_2 \ell}^2}} \ ,
\end{equation}
which is the scalar product of the vectors of normalized
weights of nodes $i_1$ and $i_2$. This quantity, which measures the
similarities between neighbourhoods of nodes, contains non-trivial
semantic information that can be used to detect synonymy relations
between tags, or to uncover ``concepts'' from social annotations
\cite{cattuto2008_cossim}. Figure \ref{fig5} shows the histograms of
pair-wise similarities between nodes in real and synthetic
co-occurrence networks. The distributions are very similar, with a skewed
behaviour and a peak for low values of the similarities.

While the data shown in Fig.s~\ref{fig3} and~\ref{fig4} correspond to
a particular example of underlying network (a Watts-Strogatz network,
see~\cite{Watts:1998}) taken as a cartoon
for the semantic space, we have also investigated
the dependence of the synthetic network properties on the structure of
the semantic space and on the other parameters, such as $n_{RW}$ or
the distribution of the random walk lengths. Interestingly, we find an
overall extremely robust behavior for the diverse synthetic networks,
showing that the proposed mechanism reproduces the empirical data
without any need for strong hypothesis on the semantic space
structure. The only general constraints we can fix on our proposed
mechanism are the existence of an underlying semantic graph with a
small diameter and a finite average degree (random walks on a fully
connected graph would not work, for instance) and a broad distribution
of post lengths. This lack of strong constraints on the precise structure
of the underlying semantic network is actually a remarkable feature of the
proposed mechanism. The details of the underlying network will unavoidably depend
on the context, namely on the specific choice of the central tag $t^*$, and
the robustness of the generative model matches the robustness
of the features observed in co-occurrence networks from real systems.
Of course, given an empiric co-occurrence network,
a careful simultaneous fitting procedure of the various observables would be needed
to choose the most general class of semantic network structures 
that generate that specific network by means of the mechanism introduced here.
This delicate issue goes beyond the goal of this paper, and also
raises the open question of the definition of the minimal set of
statistical observables needed to specify a graph \cite{Mahadevan:2006}.


\section{Conclusions}

Investigating the interplay of human and technological factors in
user-driven systems is crucial to understand the evolution and the
potential impact these techno-social systems will have on our
societies. Here we have shown that sophisticated features of the information
networks stemming from social annotations can be captured by regarding
the process of social annotation as a collective exploration of a
semantic space, modeled as a graph, by means of a series of random
walks. The proposed generative mechanism naturally yields an explanation
for the Heaps' law observed for the growth of tag vocabularies.
The properties of the co-occurrence networks generated by this mechanism
are robust with respect to the details of the underlying graph,
provided it has a small
diameter and a small average degree. This mirrors the robustness of
the stylized facts observed in the experimental data, across different systems.

Networks of resources, users, and metadata such as tags
have become a central collective artifact of the information society.
These networks expose aspects of semantics and of human dynamics,
and are situated at the core of innovative applications.
Because of their novelty, research about their structure and evolution
has been mostly confined to applicative contexts.
The results presented here are a definite step towards a fundamental
understanding of user-driven information networks that can prompt interesting
developments,
as they involve the application of recently-developed tools from complex
networks theory to this new domain. 
An open problem, for instance, is the generalization of our modeling approach
to the case of the full hyper-graph of social annotations, of which the co-occurrence network
is a projection. Moreover, user-driven information networks lend
themselves to the investigation of the interplay between social behavior and semantics,
with theoretical and applicative outcomes such as node ranking (i.e., for search
and recommendation), detection of non-social behavior (such as spam),
and the development of algorithms to learn semantic relations
from large-scale dataset of social annotations.

\begin{acknowledgments}
The authors wish to thank A. Capocci, H. Hilhorst and V.D.P. Servedio
for many interesting discussions and suggestions. This research has
been partly supported by the TAGora project funded by the Future and
Emerging Technologies program (IST-FET) of the European Commission
under the contract IST-34721.
\end{acknowledgments}

\end{document}